\documentclass{jaa}
\usepackage{natbib}
\usepackage{graphicx}

\begin{document}\sloppy

\title{Results from AstroSAT LAXPC Observations of Hercules X-1 (Her X-1)}


\author{Leahy, D.A.\textsuperscript{1,*}, Chen, Y.\textsuperscript{1}}
\affilOne{\textsuperscript{1}Department of Physics and Astronomy, University of Calgary, Calgary, Canada.\\}


\twocolumn[{

\maketitle

\corres{leahy@ucalgary.ca}

\msinfo{31 Oct 2020}{}

\begin{abstract}
The  Large Area Proportional Counter (LAXPC) instruments onboard the AstroSat Observatory
has observed the X-ray binary system Her X-1 during AstroSAT observing sessions AO2, AO3 and TO2.
These include observations while Her X-1 is in different stages of its
35 day cycle:  Low State, Turn-on to Main High State, peak of Main High State and early decline of Main High State. 
These observations also include a number of dips and one egress of neutron star eclipse. 
Here we present light curves and softness ratio analysis for these observations and discuss new features of the spectral 
changes with 35-day phase and orbital phase. We find a new phenomenon for dips during Main High state:
About half of the dips show constant softness ratio as count rate decreases, which has not been seen before, 
and could be caused by highly ionized matter or very dense cold matter.
The other half of the dips show the normal decrease of softness ratio as count rate decreases. 
 These are caused by cold matter absorption and were previously known. 
\end{abstract}


}]


\doinum{12.3456/s78910-011-012-3}
\artcitid{\#\#\#\#}
\volnum{000}
\year{0000}
\pgrange{1--}
\setcounter{page}{1}
\lp{8}

\section{Introduction}
Her X-1/HZ Her is a persistent X-ray pulsar which continues to give new information about X-ray binary astrophysics. 
E.g. Leahy \& Wang (2020) measurements the variations in the 35-day cycle
using Swift/BAT and RXTE/ASM long-term monitoring data. 
Leahy(2015) found a large-scale electron scattering corona using eclipses of the neutron star.
Accurate measurement of the companion radius using eclipses was carried out by Leahy \& Abdallah (2014). 
 The pulse period  is being tracked with Swift/BAT  (Klochkov et al. 2009). 
 Emission lines from the Her X-1 accretion disk (Ji et al. 2009) have been analyzed to learn about the state of the disk atmosphere.

This binary, with neutron star (Her X-1) and companion (HZ Her), has masses $\simeq$1.5 $M_{\odot}$ and
 $\simeq$2.2 $M_{\odot}$, respectively, measured by Leahy \& Abdallah (2014) and Reynolds et al.
(1997).
Her X-1/HZ Her is a strong emitter at optical, ultraviolet, EUV, and X-rays, which enables studies of the binary in detail.
For example, the 1.7 day optical light curve of the companion
shows there is a Roche-lobe filling accretion disk which precesses (Gerend \& Boynton 1976). 
The inner disk and from the irradiated surface of the companion star emit in the EUV (Leahy \&
Marshall 1999; Leahy et al. 2000; Leahy 2003).

Mass accretion onto the neutron star generates  hard X-rays ($>$1 keV). 
 The X-ray pulsations are determined by the hot spot geometry and by gravitational light-bending (Leahy 2004a, 2004b). 
The soft X-ray pulsations were shown to be reprocessing by the inner disk (McCray et al.1982).
The precessing accretion disk causes the 35-day flux cycle in flux and also causes the pulse shape changes with 35-day phase 
 (Scott et al. 2000). 
The accretion disk and the cycle were measured and modeled by Shakura et al. (1998),
Scott \& Leahy (1999), Leahy (2002, 2004c), and Leahy \& Igna (2011). A main result is that
the neutron star is directly seen in Main High State, partly obscured by the disk during Short High State, and greatly obscured 
during Low State. 

The accretion stream causes the well-known absorption dips (Igna \& Leahy 2011, 2012).
The lightcurve and spectral absorption for eclipse ingresses and egresses are determined by the companion star atmosphere
 (Day et al. 1988; Leahy \& Yoshida 1995). 
 During Low State and Short High State, reflected X-rays from the companion star are detected (Abdallah \& Leahy 2015). 
For Low State and eclipses a low level of X-rays is measured (Choi et al. 1994; Leahy 1995a), which is likely
scattered by extended plasma in the binary. 
Overall, Her X-1 is well measured and show patterns which are explained by the geometry of
the accretion disk, the accretion stream from HZ Her, and a large scale corona.

India’s first space-borne astronomy observatory is AstroSat (Singh et al. 2014). 
The mission was conceived starting in 1996, built over the two decades, then launched in September 2015.
AstroSat has four science instruments that are co-aligned and provide simultaneous observations over a wide energy range. 
The energy bands are: hard X-rays with the  Large Area Proportional Counter
(LAXPC) and Cadmium-Zinc-Telluride Imager (CZTI) instruments; soft x-rays with the Soft X-ray Telescope (SXT); and
optical, near and far-ultraviolet (NUV and FUV) with the UltraViolet Imaging Telescopes (UVIT).
 LAXPC is a large area proportional counter, sensitive to the 3–100 keV band, and is described in Yadav et al. (2016)
and Antia et al (2017). 
CZTI is a coded mask imager in the 25–150 keV band, and is described in Bhalerao et al. (2017).
SXT covers the energy ranges 0.3–8 keV and is described in Singh et al. (2017).
UVIT and its calibration are described in Tandon et al. (2017), Postma et al. (2011).

The 35-day cycle has been characterized previously and our general goal is to improve our understanding of the system
geometry which causes the 35-day cycle. The notation for the states is summarized in 
Scott \& Leahy (1999): Turn-On, Main High, Low State, Short High and Low State, in  time order.
The Turn-On is defined as 35-day phase 0, and is marked by the time when the fast rise to Main High
reaches 20\% of the peak flux of Main High. The Main High and Short High are the two time intervals when
the X-ray flux is bright, and the two Low States are when the X-ray flux is faint. 
The durations of the states in 35-day phase were best measured by Leahy \& Igna (2011) using the entire set
of RXTE/PCA observations of Her X-1. They are (revised from Scott \& Leahy 1999) 
Main High: 0-0.30, Low State 1: 0.30-0.57, Short High: 0.57-0.75,
and Low State 2: 0.75-1.0. The length of the 35-day cycle is somewhat variable and the boundaries between states are 
also variable. The most comprehensive characterization of the variability of the shape and timing of the 35-day cycle is
given by Leahy \& Wang (2020). 

To understand Her X-1 better, and the states during the 35-day cycle, 
we analyze AstroSat LAXPC observations of Her X-1 in the current work. 
In Section 2, we describe the observations and the resulting Her X-1 light curves.
In Section 3, we carry out a softness ratio analysis and discuss the results. We conclude in Section 4 with a short summary.


\begin{figure}[!t]
\includegraphics[width=.99\columnwidth]{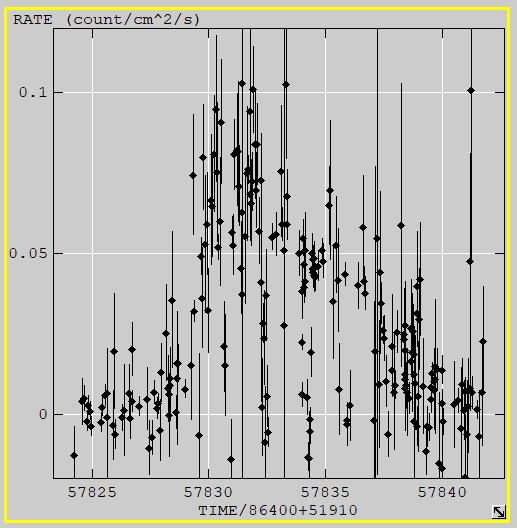}
\caption{Swift/BAT lightcurve of Her X-1 around the time of the TO2 AstroSAT observation. 
The x-axis is in MJD,  converted from Swift mission time. 
The AstroSAT Her X-1 observation for AO2 covered the time interval MJD57827.68
to  MJD57828.97, thus covering Low State and early Turn-on to Main High State. 
 }\label{fig1}
\end{figure}

\begin{figure}[!t]
\includegraphics[width=.99\columnwidth]{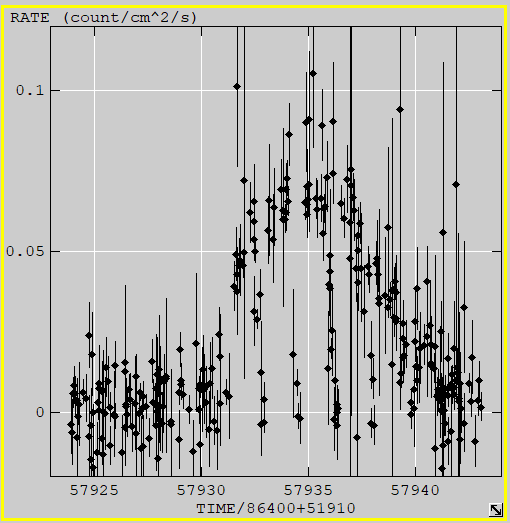}
\caption{Swift/BAT lightcurve of Her X-1 around the time of the AO3 AstroSAT observation. 
 The x-axis is in MJD, converted from Swift mission time. 
The AstroSAT observation for AO3 covered the time interval MJD57933.53 to MJD57934.27, 
thus covering the peak of Main High State. 
}\label{fig2}
\end{figure}

\section{Observations and Data Analysis}

The LAXPC is one of the main instruments on AstroSat and consists of 3 separate proportional counter detectors, each with
geometric area of 3600 cm$^2$ (Antia et al. 2017) 
and time resolution of 10 microseconds. 
The description of the LAXPC and its calibration is given by Antia et al. (2017). 
The observations of Her X-1 with LAXPC were carried out in standard event mode
as part of three separate proposals for AstroSat observing sessions AO2, AO3 and TO2.
The observation dates were Mar. 15-16, 2017 for AO2, Jun. 29-30, 2017 for AO3
and  Sep. 20-21, 2017 forTO2. For these observations only units 1 and 2 of the 3 LAXPC detectors
produced useful data. Hereafter, we label these LAXPC detectors 1 and 2 as LX10 and LX20, respectively.

\begin{figure}[!t]
\includegraphics[width=.99\columnwidth]{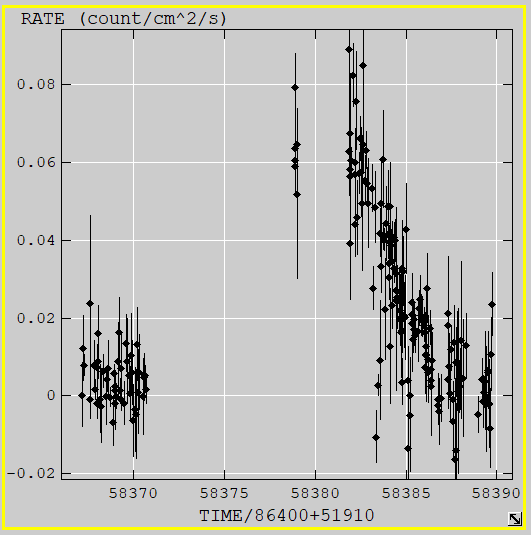}
\caption{Swift/BAT lightcurve of Her X-1 around the time of the TO2 AstroSAT observation. The x-axis is in MJD, 
converted from Swift mission time. 
The AstroSAT Her X-1 observation for TO2 covered the time interval MJD58381.83
to  MJD58382.82, thus covering the peak and start of decline of Main High State.}\label{fig3}
\end{figure}

 \begin{table*}
 \begin{center}
 \caption{LAXPC Observations of Her X-1\label{tbl-data}}
 \begin{tabular}{crrrrrr}
 \hline
Observation & MJD (start) & exposure &  $\phi_{orb,1}^{a}$ & $\phi_{orb,2}^{a}$ & $\phi_{35d,1}^{a}$   & $\phi_{35d,2}^{a}$ \\
\hline
AO2 & 57827.68 & 51523 s &  0.09 & 0.87 & 0.974 &  0.012 \\
AO3 & 57933.53 & 26786 s & 0.36 & 0.80 & 0.052 &  0.073 \\
TO2 & 58381.83 & 36380 s & 0.04 & 0.63 & 0.168 &  0.197 \\ 
 \hline
 \end{tabular}
 \tablenotes{a: 1 and 2 indicate observation start and end for orbital phase, $\phi_{orb}$, and 35-day phase, $\phi_{35d}$.}
 \end{center}
 \end{table*}
 
To determine the part of the 35-day cycle that Her X-1 was in at the time of the AstroSat LAXPC observations,
we used publically available long-term monitoring lightcurves.
The data  on Her X-1 from the Swift/BAT Hard X-ray Transient Monitor data archive Krimm et al. (2013) was 
downloaded and analyzed.
The resulting lightcurves for the $\sim$10 day period around the times of the LAXPC observations are shown
in Figures 1, 2 and 3. For TO2, there is a data gap in the Swift/BAT light curve where there is no data between
MJD58371 and 58378, so we show a $\sim$20 day period.
Comparison with lightcurves for longer timescales (35 days or more) shows that the peak
of the plotted Swift/BAT lightcurves is peak of Main High State at 35-day phase 0.10 (see Fig. 2 of 
Leahy \& Igna 2011 for the combined view of the Main High and Short High from all RXTE/PCA observations). 
Thus we see that the three AstroSat observations occured roughly around Main High state.

The Swift/BAT data (Fig. 1) shows that the AO2 observation occurred during Low State prior to Main High and the
early part of the Turn-On-to-Main-High period. For AO3, the Swift/BAT data (Fig. 2) shows the AstroSat data 
occurs during peak of Main High or just before peak but well after Turn-On. 
 For TO2, the Swift/BAT data (Fig. 3) shows the AstroSat data 
occurs during Main High, likely after peak, and includes the early part of the Main High decline
that starts at $\simeq$MJD58382.5 and continues down to Low State at $\simeq$MJD58387. 

The Turn-On (hereafter TO) times of Her X-1 for each Main High state can be measured 
 to better than 1 day using the SWIFT/BAT lightcurves (when there is data), which results in
an error in 35-day phase of $\lesssim$0.03. 
We used the TO of Her X-1 measured from the Swift/BAT lightcurves (or nearest TO for the AstroSat TO2 observation), 
to set 35-day phase 0.0 for the AstroSat LAXPC observations and calculated 35-day phase using a
mean length of the 35-day cycle of  34.9 days (Leahy \& Wang 2020). 
We determined orbital phase using the ephemeris of Staubert et al. (2009). 
Table~\ref{tbl-data} summarizes the observations, including MDJ, orbital phase and 35-day phase.

The data was downloaded from the AstroSat Data Archive at the AstroSat Science Support Cell
(hereafter ASSC) website. 
The data analysis was carried using the LAXPC software package from the ASSC.
Data segments corresponding to the time intervals during earth occultation of source and satellite
passage through the South Atlantic Anomaly (SAA) region were removed before the creation of light curves
and spectra.
 We used the LAXPC data from all layers combined.



\subsection{Light-curve Analysis}

\begin{figure}[!t]
\includegraphics[width=1.1\columnwidth]{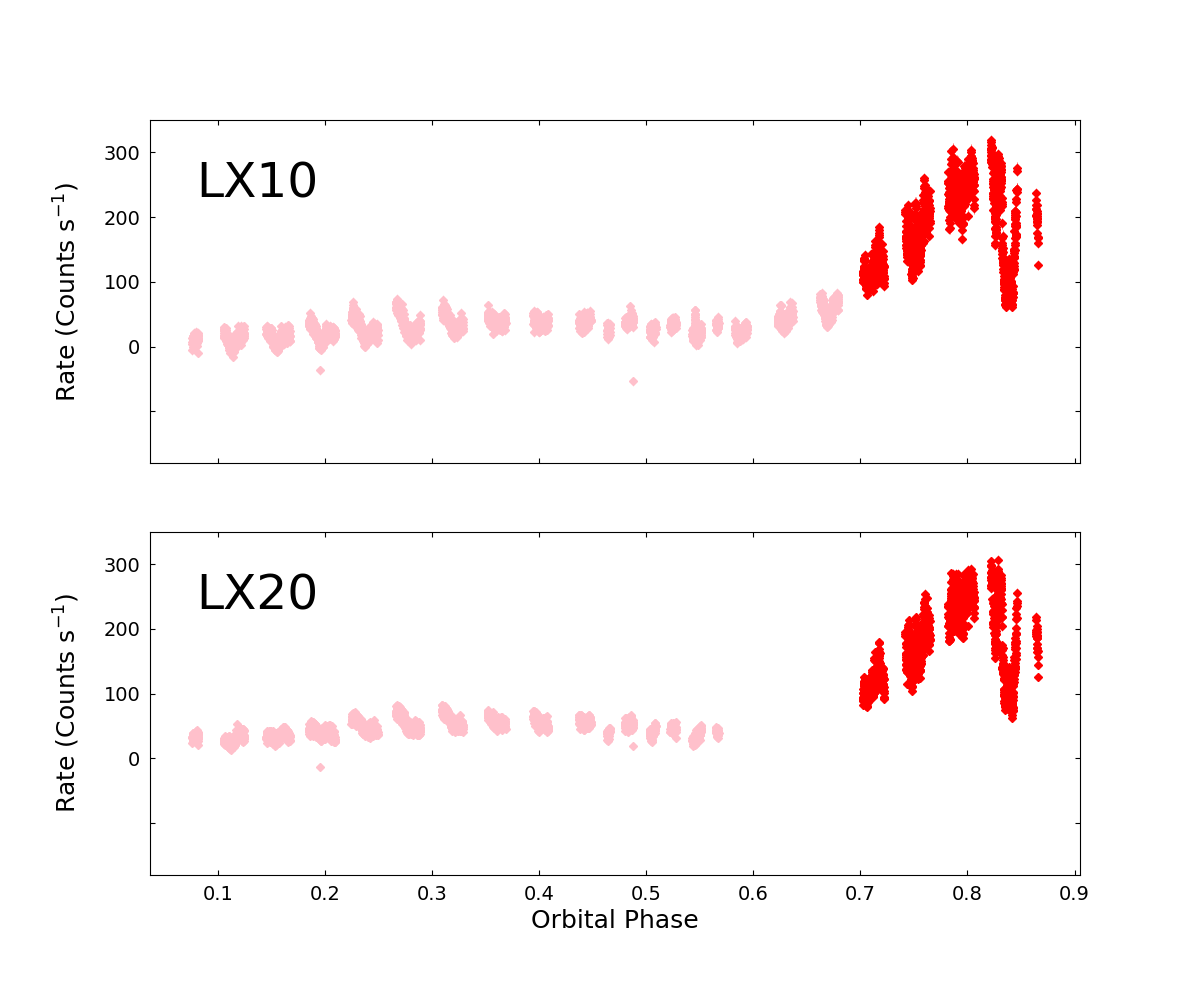}
\caption{AstroSAT LAXPC 3-80 keV lightcurve of Her X-1 for the AO2 AstroSAT observation from LAXPC unit 1 (top, labelled LX10)
and from LAXPC unit 2 (bottom, labelled LX20).
The points are color-coded by 35-day state: pink for Low State and red for Turn-on.
Fig.1 (top panel) of  Leahy \& Chen (2019) shows the SXT lightcurve. }\label{fig4}
\end{figure}

Background subtraced light curves were created for each of AO2, AO3 and TO2 in the energy bands 3-5 keV, 5-9 keV,
9-20 keV and total band of 3-80 keV. The 3-80 keV light curves extracted from AO2 LAXPC unit 1 and LAXPC unit 2 are
shown in Figure 4. Here we plot vs. orbital phase because of the importance of timing of eclipses which block direct 
X-rays from the neutron star. For AO2, the observation starts after end of eclipse at orbital phase 0.065 and ends
prior to start of the following eclipse at orbital phase 0.935. 
The temporary drop in count rate seen at orbital phase $\simeq$0.83 is identified as an absorption dip.
Points corresponding to Low State are marked in pink in this plot and points during the Turn-On phase are marked in red. 

\begin{figure}[!t]
\includegraphics[width=1.0\columnwidth]{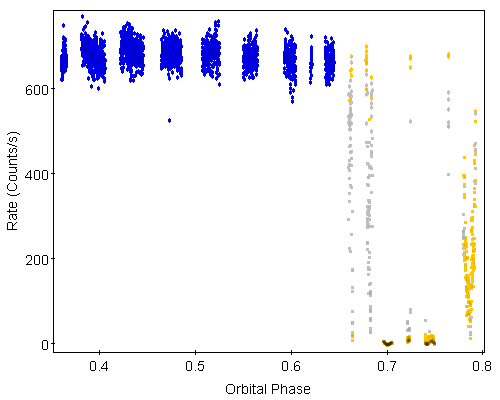}
\caption{AstroSAT LAXPC 3-80 keV lightcurve of Her X-1 for the AO3 AstroSAT observation from LAXPC 
unit 2.
The points are color-coded by 35-day state: blue for Main High State (MH) and orange and grey for MH with Dips
(see text for description).
 The  lightcurve for LAXPC1 looks similar, as in Figure 4.
Fig.1 (bottom panel) of  Leahy \& Chen (2019) shows the SXT lightcurve. }\label{fig5}
\end{figure}

\begin{figure}[!t]
\includegraphics[width=1.1\columnwidth]{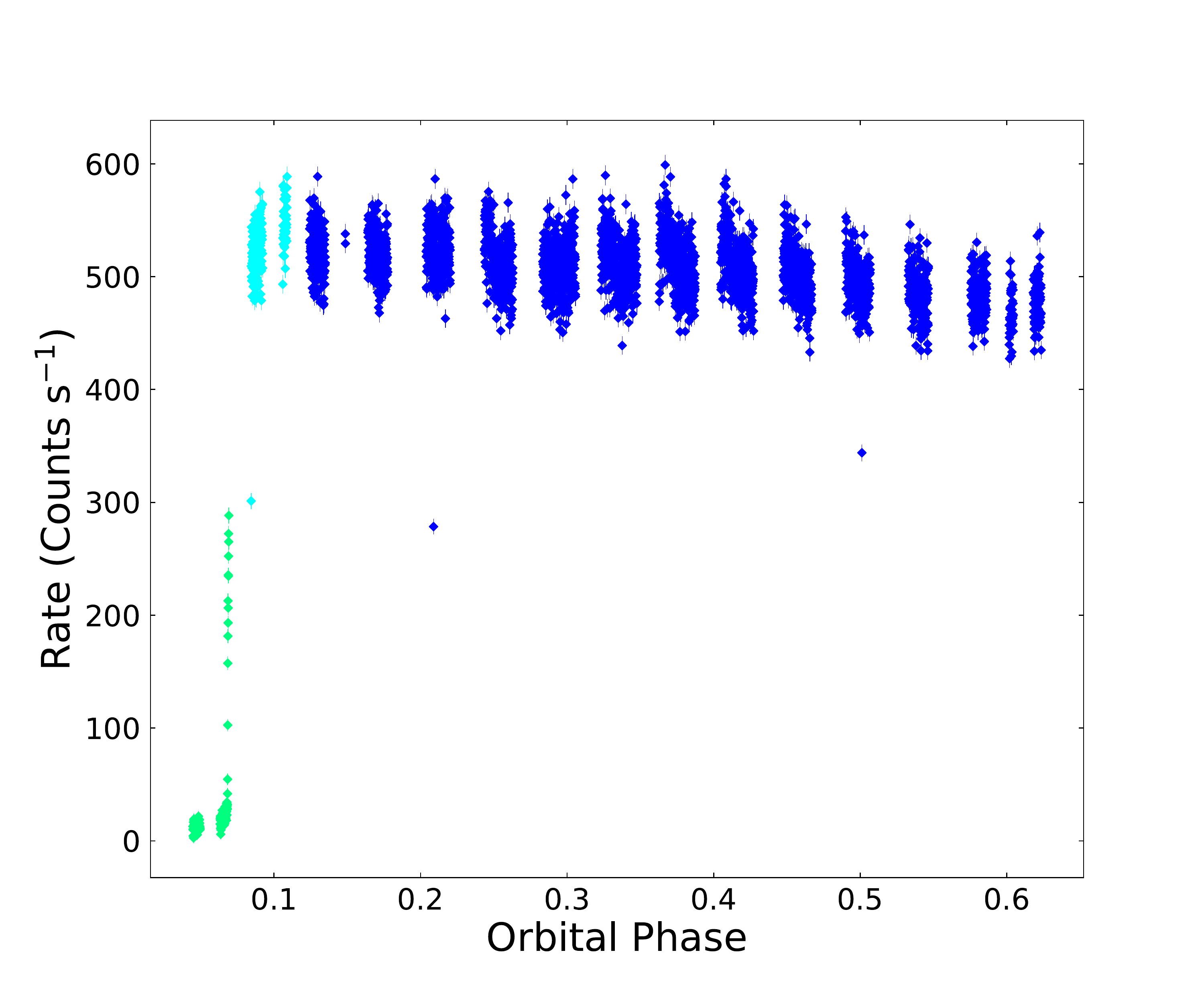}
\caption{AstroSAT LAXPC 3-80 keV lightcurve of Her X-1 for the TO2 AstroSAT observation for LAXPC unit 2.
The points are color-coded by 35-day state: blue for Main High, green for Eclipse Egress and light blue
for  the 2 hours  immediately after Main High eclipse egress.
The  lightcurve for LAXPC1 looks similar, as in Figure 4.
\ }\label{fig6}
\end{figure}

The 3-80 keV light curves extracted from AO3 for LAXPC units 1 and 2 are
shown in Figure 5, plotted vs. orbital phase.  
For AO3, the observation includes no eclipses but exhibits strong and numerous dips between orbital phase
0.65 and the end of the observation at orbital phase 0.80. 
Points during Main High State are marked in blue and points during the dips are marked in orange.  

For TO2, the 3-80 keV light curves were for LAXPC units 1 and 2, and are very similar. We show the
LAXPC unit 2 light curve in Figure 6, plotted vs. orbital phase.  
The TO2 observation includes an eclipses and eclipse egress at the beginning (orbital phase 0.05 to 0.08).
The remainder of the observation is Main High State.
Points during eclipse and egress are marked in green and points during Main High State are marked in blue.
In order to test whether the first 2 hours after eclipse are different than the rest of Main High, we analyzed
the X-ray colors later for that time period separately, and mark those 
points in light blue in Fig. 6.  

\begin{figure*}
\centering\includegraphics[width=1.7\columnwidth]{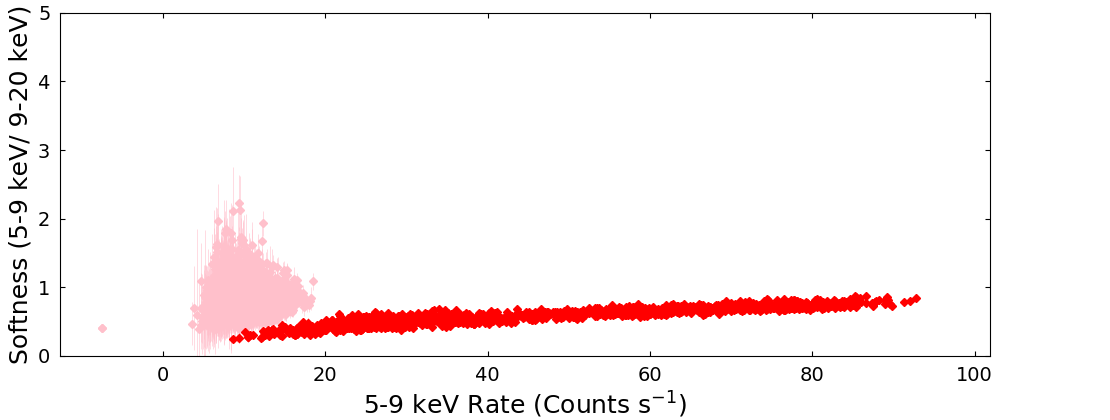}
\centering\includegraphics[width=1.55\columnwidth]{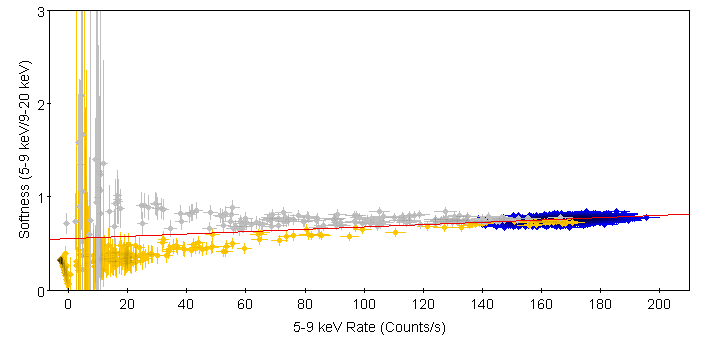}
\centering\includegraphics[width=1.65\columnwidth]{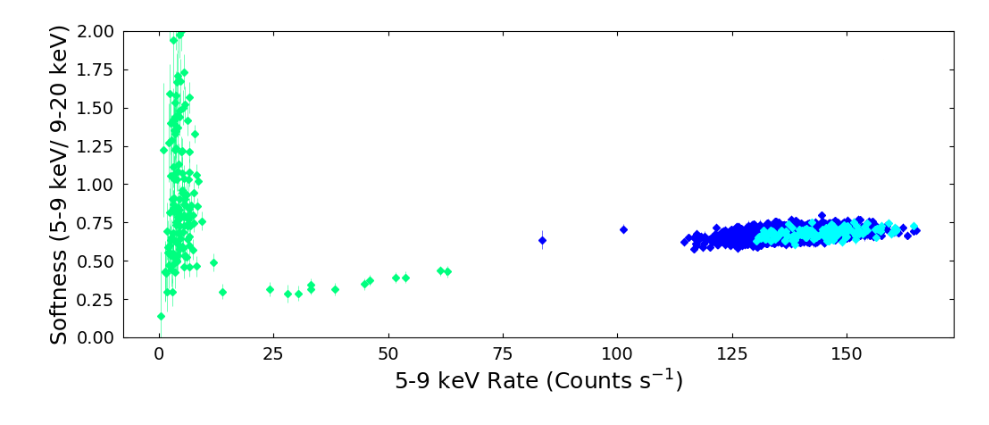}
\caption{AstroSAT LAXPC softness ratio vs count rate for the AO2 (top), AO3 (middle) and TO2 (bottom)
AstroSAT observations, shown for LAXPC unit 2.
The points are color-coded by 35-day state: pink for Low State and red for Turn-on (top panel),
blue for Main High, orange and grey for Dips (middle panel, see text for description), 
green for Eclipse Egress and light blue for the 2 hours  immediately after Main High eclipse egress (bottom panel).
The red line in the middle panel is used to divide the Dips data into two parts: those with high softness ratio 
(grey) and those with low softness ratio (orange).}
\end{figure*}

\section{Results and Discussion}

Her X-1 has a spectrum (reviewed in Leahy \& Chen, 2019), which during Main High State, is characterized 
by a broken power law with flat low energy index ($\alpha\simeq 0.9$) below a break energy of $\sim$10 keV
and steeper index ($\alpha\simeq 1.5$) above that energy. The power law is modified by an exponential 
cut-off with cut-off energy  $\sim$22 keV and e-folding energy  $\sim$12 keV. 
Other components include: a cyclotron absorption line at $\sim38$ keV, 
iron emission lines at 6.4 and 6.6 keV, an iron L emission complex at 0.9 keV with width $\sim0.2$ keV,
and a blackbody with kT=0.09 keV. On top of this spectrum is a partial covering absorber, with variable 
covering fraction and with absorption column density of $\sim5\times10^{23}$ cm$^{-2}$.
Leahy \& Chen (2019) found from AstroSat SXT observations that the presence of an additional partially ionized 
absorber was strongly indicated by the 2-8 keV SXT spectrum during Main High State.

Spectral changes in Her X-1 over the 35-day cycle or over the orbital period are primarily driven by
the system geometry. As reviewed in Leahy \& Igna (2011) (see also Leahy, 1995c)  the changes in spectrum
can be summarized by changes in softness ratio. The changes in system geometry include eclipse of 
neutron star by Her X-1 (Leahy \& Abdallah 2014), absorption dips from the accretion stream (Igna \& Leahy 2011), 
and absorption re-emission and scattering  by the accretion disk (Leahy 2002).
Here we study the time variability of Her X-1 in the LAXPC observations using softness ratio,
with detailed spectral analysis to be carried out in future work.

\subsection{Softness Ratio Analysis}

Previous analyses of the 35-day cycle and the properties of the different states 
(Turn-On, Main High, Low State, Short High, Eclipse,  Eclipse Ingress and Egress, and Dips)
were carried out by Leahy \& Igna (2011) and Leahy (1995c). The 
former used RXTE/PCA observations and the latter used GINGA observations of Her X-1.
X-ray color (softness ratio) vs. intensity plots were made for the different states, showing 
characteristic features of the different states. E.g. Fig. 4 in  Leahy (1995c) shows eclipse and eclipse
egress in Main High state, and dips during Main High state.
Fig. 6 in  Leahy (1995c)  shows a summary plot with all of the points for different states labelled by state,
illustrating the utility of a color-intensity diagram as a diagnostic for the Her X-1 states.

Here we have created color-intensity diagrams for the three observations (AO2, AO3 and TO2) using 
5 to 9 keV count rate for intensity and (5 to 9 keV)/(9-20 keV) softness ratio for color. 
The results are color-coded by state as noted above in the light curve plots of Figures 4, 5 and 6.
Figure 7 shows these color-intensity plots. 

The Low State (pink), the deepest part of dips (orange points at count rate at $<$5 counts/s ) 
and eclipse (green points at count rate $\lesssim$10 counts/s) all have the same softness ratio (1.0 within errors).
With the current background subtraction for LAXPC, the Low State is significantly brighter than 
deep dips or eclipse. 

Next we compare the Turn-On phase (red), dip ingress and egress (orange and grey points at count rate at $\gtrsim$10 counts/s )
and  eclipse egress (green points at count rate $\gtrsim$10 counts/s). 
The softness ratio increases from 0.25 to 0.8 for Turn-On (red points in Fig. 7), with final value of 0.8 
the same as for the Main High state (blue points in the lower two panels of Fig. 7). 

Dip ingress and egress (the time period between onset of dip and when the count rate reaches its
lowest value during dip) shows two different branches in softness ratio (Fig. 7 middle panel, orange and
grey points).
The lower branch changes in softness ratio from 0.8 (normal Main High softness ratio) down to 0.25 when the
count rate reaches its minimum ($\sim$5 counts/s).
The upper branch is, in contrast, nearly constant in softness ratio at 0.8 (normal Main High softness ratio)
while the count rate decreases from its  maximum ($\sim$170 counts/s) to its  minimum ($\sim$5 counts/s).
Here, we chose the line of softness ratio (SR) given by $SR=R(5-9)/800+0.5$ with $R(5-9)$ the 5-9 keV
count rate, to divide the upper branch set from the lower branch set of dip points. This dividing line is
shown in the middle panel of Fig. 7 by the red line. We note that at
high 5-9 keV count rate (above $\sim$100 counts/s) the line is not able to separate the two branches
because they overlap too closely. 
At low count rates, the softness ratio errors are too large to separate the two branches.

Referring back to the light curve for MH and MH with Dips (Fig. 5), it is seen that there are 3 clear
dips during the observation.
The two dips with constant softness ratio (grey points) are the sharp dips at orbital phase $\simeq$0.66 and 
$\simeq$0.68. The dip with softness ratio decreasing with decreasing count rate is the wider and slower
dip which occurs from orbital phase $\simeq$0.78 to $\simeq$0.80. 

Eclipse egress (green points in lower panel) starts at softness ratio $\sim$0.25 and rises smoothly with count rate 
to the normal 0.8 value for Main High state. Fig. 6 shows that the final part of
egress from 5-9 keV count rate of 70 to 125 counts/s is missing in the AstroSat data. 
The two blue points at count rate $\sim$80 and 105 are not part of egress, rather small short dips, as seen from Fig. 6.

The two hours of data just after eclipse egress (see Fig. 6) are separately plotted in Fig. 7 as the light
blue points. It is seen that they cannot be distinguished from the rest of Main High state. 
This indicates that the effects of eclipse are finished within a short time ($\lesssim$50 min) after egress.  

\subsection{Implications}

The light curves obtained during AO2, AO3 and TO2 exhibit the various known states of Her X-1 for the parts
of the 35-day cycle that were observed. The Low State, Turn-On, Main High state, Main High dips and one eclipse
and its egress were well measured.
A color-intensity diagram analysis, similar to that carried out for GINGA data by Leahy (1995c) was carried out.
The AstroSat LAXPC color-intensity diagram has features that agree with those from the color-intensity diagram
for GINGA. 
However the new observations and analysis with AstroSat have smaller errors than those from the GINGA
analysis.

One new feature that we have found comes from the extensive dipping period during peak of Main High state during
the AstroSat AO3 observation. For the first time, we detect two types of dips which can be separated by the
count-rate dependence of their softness ratios. 

As seen in the middle panel of Fig. 7, we separated the dip data into two sets:  dips which show a decrease softness ratio as the count rate decreases (orange points); and dips which show a constant
softness ratio as the count rate decreases (grey points).
The first type of dips behave the same way as eclipse egress or as Turn-On.  Previous spectral analyses of
eclipse egress (Day et al. 1988,  Leahy \& Yoshida 1995) have shown 
this is caused by cold matter absorption changing both intensity and softness ratio.
Similarly analysis of Turn-On has shown the changes are caused by cold matter absorption (Kuster et al. 2005).

The newly discovered dips have nearly constant softness ratio as the count rate decreases. These dips could be caused
by highly ionized matter blocking the X-rays to the neutron star, so that the spectrum does not change 
as count rate decreases. Alternately, they could be caused by partial coverage by very dense matter
(column density $\gtrsim 10^{24}$ g~cm$^{-2}$), which does not change the spectrum as count rate decreases,
but requires a sharp physical edge to the matter. 




\section{Conclusion}

For the current work, we have analyzed observations of Her X-1 from the Large Area Proportional Counter (LAXPC) instruments onboard the AstroSat Observatory.
The observing sessions  include observations while Her X-1 is in different stages of its
35 day cycle:  Low State, Turn-on to Main High State, peak of Main High State and early decline of Main High State. 
The observations also include a number of dips and one egress of neutron star eclipse. 

We have presented light curves and a softness ratio analysis. Many results confirm previous findings for 
Her X-1 on the behaviour of the different states. 
A fraction $\sim$1/2 of the dips show decreasing softness ratio as count rate decreases. 
 These are the previously known type of dips and are caused by cold matter absorption. 
 However, a new phenomenon is found for dips during Main High state:
 the other fraction $\sim$1/2 of dips are new: they show constant softness ratio as count rate decreases.
These dips could be caused by highly ionized matter or by very dense cold matter with a sharp edge.
 
 Future work will focus on verifying and understanding the new type of dips by carrying out a spectral analysis 
 with both LAXPC and SXT instruments. New observations are planned for Short High state to characterize it better,
 including a search for dips to test if Short High dips behave in the same way as Main High dips.

\section*{Acknowledgements}
This project is undertaken with the financial support of the
Canadian Space Agency and of the Natural Sciences and
Engineering Research Council of Canada.
This publication uses data from the AstroSat
mission of the Indian Space Research Institute
(ISRO), archived at the Indian Space Science Data Center
(ISSDC). LAXPC data were processed by the Payload Operation Centre at TIFR, Mumbai.
\vspace{-1em}


\end{document}